\documentclass[preprint2]{aastex}
\usepackage{spr-astr-addons}
\usepackage{graphicx}
\begin{document}

\title{Jet confinement by magneto-torsional oscillations
}


\author{Gennady Bisnovatyi-Kogan
}


\affil{Space Research Institute, Profsoyuznaya 84/32, Moscow
117997, Russia\\
              Tel.: +7-495-3334588\\
              Fax:  +7-495-3334588\\
              \email{gkogan@iki.rssi.ru}           
           \and
}

\date{Received: date / Accepted: date}

\maketitle

\begin{abstract}
Many quasars and active galactic nuclei (AGN) appear in radio,
 optical, and X-ray maps,
as a bright nuclear sources from which emerge single or double long,
thin jets. When observed with high angular
resolution these jets show structure with bright knots separated by
relatively dark regions. Nonthermal nature of
a jet radiation is well explained as the synchrotron radiation
of the relativistic electrons in an ordered magnetic field.
 We consider
magnetic collimation, connected with torsional oscillations of a
cylinder with elongated magnetic field, and periodically
distributed initial rotation around the cylinder axis. The
stabilizing azimuthal magnetic field is created here by torsional
oscillations, where charge separation is not necessary. Approximate
simplified model is developed. Ordinary differential equation is
derived, and solved numerically, what gives a possibility to
estimate quantitatively the range of parameters where jets may be
stabilized by torsional oscillations.
\end{abstract}

\section{Introduction}
\label{intro}
Objects of different scale and nature in the universe:
from young and very old stars to galactic nuclei show existence of
collimated outbursts - jets. Geometrical sizes of jets lay between
parsecs and megaparsecs. The origin of jets is not well understood
and only several qualitative mechanisms are proposed which are not
justified by calculations. Theory of jets must give answers to three
main questions: how jets are formed? how are they stabilized? how do
they radiate? The last question is related to the problem of the
origin if relativistic particles in outbursts from AGN, where
synchrotron emission is observed. Relativistic particles, ejected
from the central machine rapidly loose their energy so the problem
arises of particle acceleration inside the jet, see review of
Bisnovatyi-Kogan (1993).

 It is convenient sometimes to investigate
jets in a simple model of infinitely long circular cylinder,
Chandrasekhar \& Fermi (1953).
 The magnetic field in the collimated jets determines its
direction, and the axial current may stabilize the jet's elongated
form at large distances from the source (e.g. in AGNs),
(Bisnovatyi-Kogan et al., 1969). When observed with high angular resolution these jets
show a structure with bright knots separated by relatively dark
regions (Thomson et al., 1993).
 High percentages of polarization, sometimes
exceeding 50\%, indicate the nonthermal nature of the radiation,
explained as synchrotron emission of the relativistic
electrons in an ordered magnetic field.

 Here we consider stabilization of a jet by magnetohydrodynamic mechanism,
 connected with torsional oscillations.
 We suggest that the matter in the jet is
 rotating, and different parts of the jet rotate in different directions. Such distribution of
the rotational velocity produces azimuthal magnetic field, which
prevents a disruption of the jet. The jet remains to be in a
dynamical equilibrium, when it is representing a periodical, or
quasiperiodical structure along the axis, and its radius is
oscillating with time all along the axis. The space and time period
of oscillations depend on the conditions at jet formation: the
length scale, the amplitude of the rotational velocity, and the
strength of the magnetic field. The time period of oscillations
should be obtained during construction of the dynamical model, what
also should show at which input parameters may exist a long jet,
stabilized by torsional oscillations.

\begin{figure}
\centerline { \includegraphics[width=8cm,angle=00]{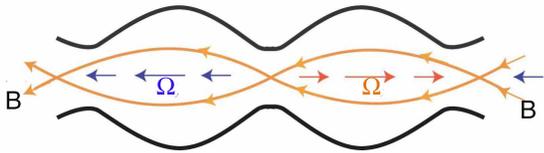}}
\caption{Jet confinement by magneto-torsional oscillations
(qualitative picture)} \label{fig1}
\end{figure}
 Here we construct a simplified model of this
phenomena, which confirms the reality of such stabilization, see
also Bisnovatyi-Kogan (2007).

\section{General picture}
\label{sec:1}
Consider a long cylinder with a magnetic field directed along its
axis (Fig.1). This cylinder will expand unlimitly under the action of
pressure and magnetic forces. The
limitation of the radius of this cylinder could be possible in
dynamic state, when the whole cylinder undergoes magneto-torsional
oscillations. Such oscillations produce toroidal field, which
prevent a radial expansion. There is a competition between the
induced toroidal field, compressing the cylinder in radial
direction, and gas pressure, together with the field along the
cylinder axis (poloidal), tending to increase its radius.During
magneto-torsional oscillations there are phases, when either
compression or expansion forces prevail, and, depending on the input
parameters, we may expect tree kinds of a behavior of such cylinder.

1. The oscillation amplitude is low, so the cylinder suffers
unlimited expansion (no confinement)

2. The oscillation amplitude is too high, so the pinch action of the
toroidal field destroys the cylinder, and leads to formation of
separated blobs.

3. The oscillation amplitude is moderate, so the cylinder survives
for an unlimited time, and its parameters (radius, density, magnetic
field etc.) change periodically, or quasi-periodically in time.
Here we try to find a simple approximate way
for obtaining a qualitative answer, and to make a rough estimation
of parameters leading to different regimes.

\section{Profiling in axially symmetric MHD equations}
\label{sec:2}

We simplify the system of MHD equations with axial symmetry,
$\frac{\partial}{\partial\phi}$=0, for the perfect gas at infinite
conductivity, written in cylindric coordinates $(r,\phi,z)$ (Landau
and Lifshits, 1982). We use for this purpose a profiling procedure.
There is no gravity in the direction of the cylinder axis ($z$), and
we approximate density by a function $\rho(t,z)$, suggesting a
uniform density along the radius. The components of the velocity and
magnetic field are approximated as

\begin{equation}
\label{eq15}
  v_r=r\, a(t,z), \quad v_\varphi=r\, \Omega(t,z),\quad v_z=0;
 \end{equation}

\begin{equation}
\label{eq16} B_r=r\, h_r(t,z), \quad B_\varphi=r\, h_\varphi(t,z),
B_z=B_z(t,z).
 \end{equation}
In this case the current components are
written as

\begin{equation}
\label{eq17}
 j_r=-\frac{cr}{4\pi}\frac{\partial h_\varphi}{\partial z},\quad
 j_\varphi=\frac{cr}{4\pi}\frac{\partial h_r}{\partial z},\quad
 j_z=\frac{ch_\varphi}{2\pi}.
 \end{equation}
After neglecting velocity $v_z$ along the axis, we should omit the
corresponding Euler equation, and the radial pressure gradient is
approximated by the linear function, what gives, when using
appropriate dimensions

\begin{equation}
\label{eq18} \frac{\partial P}{\partial r}= \lambda \frac{P}{R^2}r,
 \end{equation}
where the constant $\lambda \sim 1$ is connected with the equation
of state, $P(t,z)$ is the pressure, $R(t,z)$ is the radius of the
cylinder. In the subsequent consideration we consider an adiabatic
case with a polytropic equation of state $P=K\,\rho^\gamma$. Neglecting
the $z$ derivatives in the Poisson equation, we obtain
$\varphi_G=\pi G \rho r^2$. Substituting (\ref{eq15})-(\ref{eq18})
into the original system of the equations, we obtain for the
profiling functions the following equations

\begin{equation}
\label{eq19} \frac{\partial a}{\partial
t}+a^2-\Omega^2=\lambda\frac{P}{\rho R^2} -2\pi
G\rho+\frac{1}{4\pi\rho}\left(B_z\frac{\partial h_r}{\partial z}-2
h_\varphi^2\right),
 \end{equation}

\begin{equation}
\label{eq20} \frac{\partial \Omega}{\partial t}+2a\Omega=
\frac{1}{4\pi\rho}\left(B_z\frac{\partial h_\varphi}{\partial z}+2
h_r h_\varphi\right),
 \end{equation}

\begin{equation}
\label{eq22} \frac{\partial h_\varphi}{\partial t}=\frac{\partial
(\Omega B_z)}{\partial z} -2(a h_\varphi-\Omega h_r),
 \end{equation}

\begin{equation}
\label{eq21} \frac{\partial h_r}{\partial t}=\frac{\partial
(aB_z)}{\partial z},\quad
 \frac{\partial B_z}{\partial t}=-2 a B_z,
 \end{equation}

\begin{equation}
\label{eq24} \frac{\partial \rho}{\partial t}=-2 a \rho,
 \quad \frac{\partial R}{\partial t}=a R.
 \end{equation}
It follows from (\ref{eq21}),(\ref{eq24}) relations, representing
conservation of mass, and magnetic flux equivalent to freezing
condition

\begin{equation}
\label{eq26} \rho\, R^2=\,C_m(z), \quad B_z\, R^2=\, C_b(z), \quad
B_z=\frac{C_b(z)}{C_m(z)}\rho.
 \end{equation}
In our subsequent consideration the arbitrary functions will be
taken as constants: $C_m(z)=C_m, \,\, C_b(z)=C_b.$

It was shown by Bisnovatyi-Kogan (2007)that the approximate system
of equations describes correctly small perturbations, connected with
radial and torsional modes, in the equilibrium self-gravitating
cylinder. We expect therefore that these modes will be described
correctly in general nonlinear case in the jets where gravity is
neglected.

\section{Farther simplification: reducing the problem to ordinary differential equation}
\label{sec:4}
 For the relativistic jet we neglect gravity, without which
 the equilibrium static state of the cylinder does not exist. We need to solve numerically the
 system of nonlinear  equations (\ref{eq19})-(\ref{eq26}) to check the possibility of
 the existence of a cylinder, which radius remains finite due to torsional oscillations.
Instead we reduce the system to ordinary equations.
Supposing a constant bulk motion velocity along $z$-axis,
we consider axially symmetric jet in the
comoving coordinate frame. If the confinement is reached
due to standing magneto-torsional oscillations, there are points
along $z$-axis where rotational velocity remains zero in this
frame. Taking $\Omega=0$ in the plane $z=0$, let us consider
standing wave torsional oscillations with the space period  $z_0$ along $z$
axis. Then nodes with $\Omega=0$ are situated at
$z=\pm n \frac{z_0}{2}$, $n=0,1,2,...$. Let us write the equations,
describing the cylinder behavior in the plane $z=0$, where
$\Omega=0$. All values in this plane we denote by ($\tilde {\rm
\phantom{a}}$). We take also for simplicity $\lambda=1$. We have
than equations in the plane $z=0$ as

\begin{equation}
\label{eq40} \frac{d{\tilde a}}{d t}+\tilde a^2=\frac{K}{\tilde R^2}
+\frac{C_b}{4\pi C_m}\left(\frac{\partial h_r}{\partial
z}\right)_{z=0} -\frac{\tilde h_\varphi^2\tilde R^2}{2\pi C_m},
\end{equation}
\begin{equation}
\label{eq41} \frac{C_b}{4\pi C_m}\left(\frac{\partial
h_\varphi}{\partial z}\right)_{z=0} +\frac{\tilde h_\varphi \tilde
h_r \tilde R^2}{2\pi C_m}=0,
\end{equation}
\begin{equation}
\label{eq42} \frac{d\tilde h_r}{d t}=C_b \left(\frac{\partial
(a/R^2)}{\partial z}\right)_{z=0},
 \end{equation}
\begin{equation}
\label{eq43} \frac{d \tilde h_\varphi}{d t}=
C_b\left(\frac{\partial(\Omega/R^2)}{\partial z}\right)_{z=0}
-2\tilde a \tilde h_\varphi,
\quad
 \frac{d \tilde R}{d t}=\tilde a \tilde R,
\end{equation}
\begin{equation}
\label{eq45} \tilde \rho\, \tilde R^2=\,C_m, \quad \tilde B_z\,
\tilde R^2=\, C_b, \quad \tilde B_z=\frac{C_b}{C_m}\tilde\rho.
\end{equation}
Initial conditions for the system (\ref{eq40}) -  (\ref{eq45}) are

\begin{equation}
\label{eq46} \tilde R=R_0,\,\, \tilde\rho=\rho_0=\frac{C_m}{ R_0^2},
\,\, \tilde B_z=\frac{C_b}{ R_0^2},\,\,
 \end{equation}
$$\tilde a=\tilde h_r=\tilde
h_\varphi=0\,\, {\rm at}\,\, t=0.
$$
In (\ref{eq40}) - (\ref{eq45}) we have used relations

\begin{equation}
\label{eq47} \tilde \rho=\rho_0 \frac{R_0^2}{\tilde R^2}, \,\,
\tilde B_z=\rho_0 \frac{C_b}{C_m}\frac{ R_0^2}{\tilde R^2},
 \end{equation}
valid for any time. If the cylinder rotational velocity is
antisymmetric relative to the plane $z=0$, $\Omega=0$, and cylinder
density distribution is symmetric relative to this plane, then we
have extremum (maximum) of the azimuthal magnetic field $h_\phi$,
with $\left(\frac{\partial h_\varphi}{\partial z}\right)_{z=0}=0$,
and zero value of $\tilde h_r=0$, which reaches an extremum
(minimum) in this plane with $\left(\frac{\partial h_r}{\partial
z}\right)_{z=0}=0$. The product $a\rho$ also reaches an extremum in
the plane $z=0$, so that $\left(\frac{\partial (a/R^2)}{\partial
z}\right)_{z=0}=0$. The term with $z$ derivative in the equation
(\ref{eq43}) is not equal to zero, and changes periodically during
the torsional oscillations. We substitute approximately the
derivative $d/dz$ by the  ratio $1/z_0$, where $z_0$ is the space
period of the torsional oscillations along $z$ axis. While
$\Omega=0$ in the plane $z=0$, its derivative along $z$ is changing
periodically with an amplitude $\Omega_0$, and frequency $\omega$,
which should be found from the solution of the problem. We
approximate therefore

\begin{equation}
\label{eq48} \left(\frac{\partial(\Omega/R^2)}{\partial
z}\right)_{z=0} =\frac{\Omega_0}{z_0\tilde R^2}\cos{\omega t}.
\end{equation}
Finally, we have from (\ref{eq40}),(\ref{eq43}),(\ref{eq45}) the
following approximate system of equations, describing the non-linear
torsional oscillations of the cylinder at given $z_0$ and
$\Omega_0$.

$$
\frac{d{\tilde a}}{d t}+\tilde a^2=\frac{K}{\tilde R^2}
-\frac{\tilde h_\varphi^2 \tilde R^2}{2\pi C_m},
 $$
\begin{equation}
\label{eq49} \frac{d \tilde h_\varphi}{d t}= C_b
\frac{\Omega_0}{z_0\tilde R^2}\cos{\omega t} -2\tilde a \tilde
h_\varphi,
\quad
\frac{d \tilde R}{d t}=\tilde a \tilde R.
\end{equation}
The combination of last two equations gives

\begin{equation}
\label{eq50} \frac{d (\tilde h_\varphi R^2)}{d t}=
 \frac{C_b\Omega_0}{z_0}\cos{\omega t},
 \end{equation}
has a solution, satisfying initial condition (\ref{eq46}),
$ \tilde h_\varphi R^2=
\frac{C_b\Omega_0}{z_0\omega}\sin{\omega t},$
with account of which the first and third equations in (\ref{eq49}) are written as

\begin{equation}
\label{eq52} \tilde R\frac{d (\tilde a R)}{d t}=
 K-\left(\frac{C_b\Omega_0}{z_0\omega}\right)^2\frac{\sin^2{\omega t}}{2\pi C_m},
 \end{equation}

\section{Numerical solution}
\label{sec:5}
\begin{figure}
\includegraphics[angle=-90,width=0.45 \textwidth]{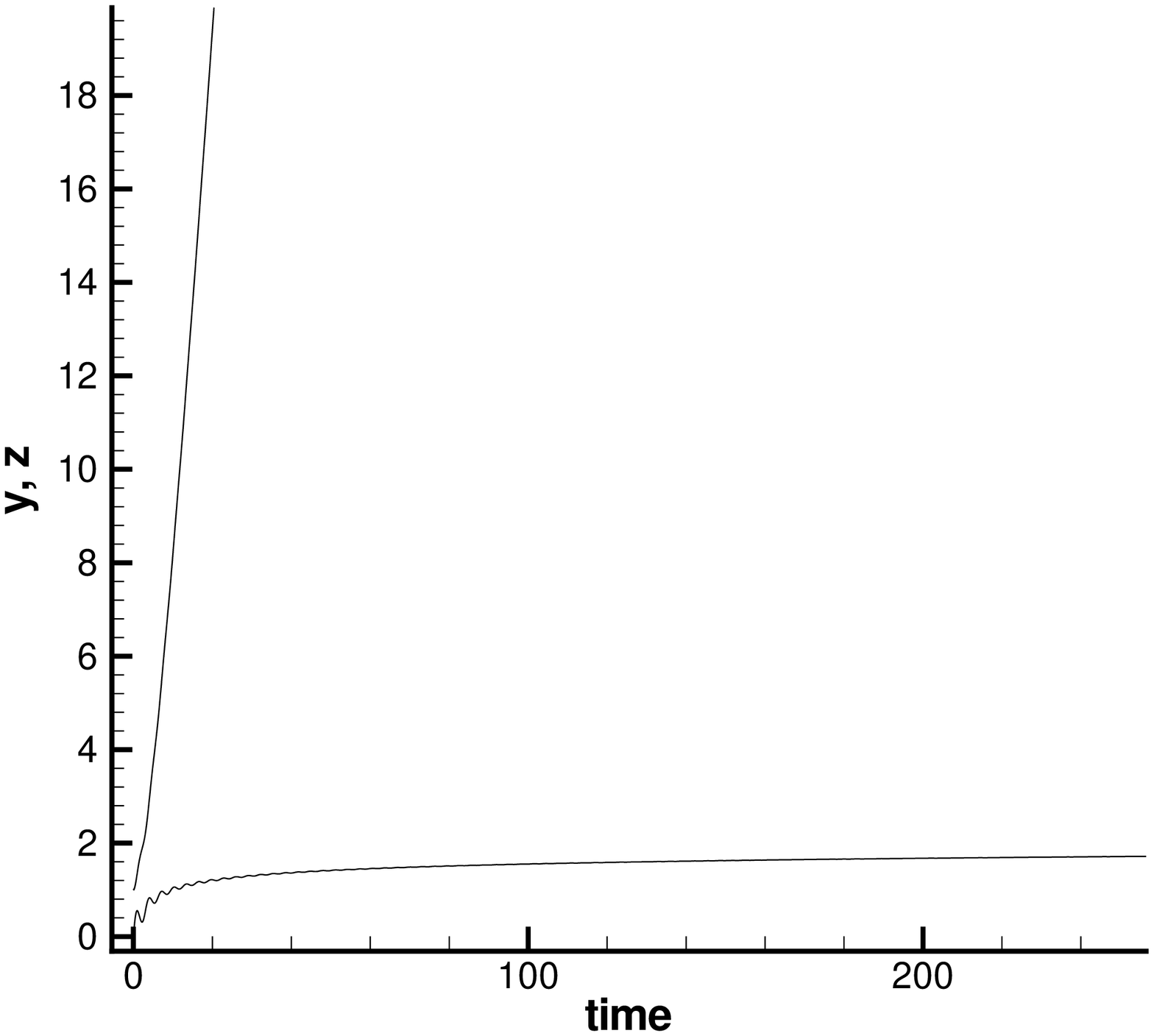}
\caption{Time dependence of non-dimensional radius $y$ (upper
curve), and non-dimensional velocity $z$ (lower curve), for
$D=1.5$.} \label{fig2}
\end{figure}

\begin{figure}
\includegraphics[angle=-90,width=0.45 \textwidth]{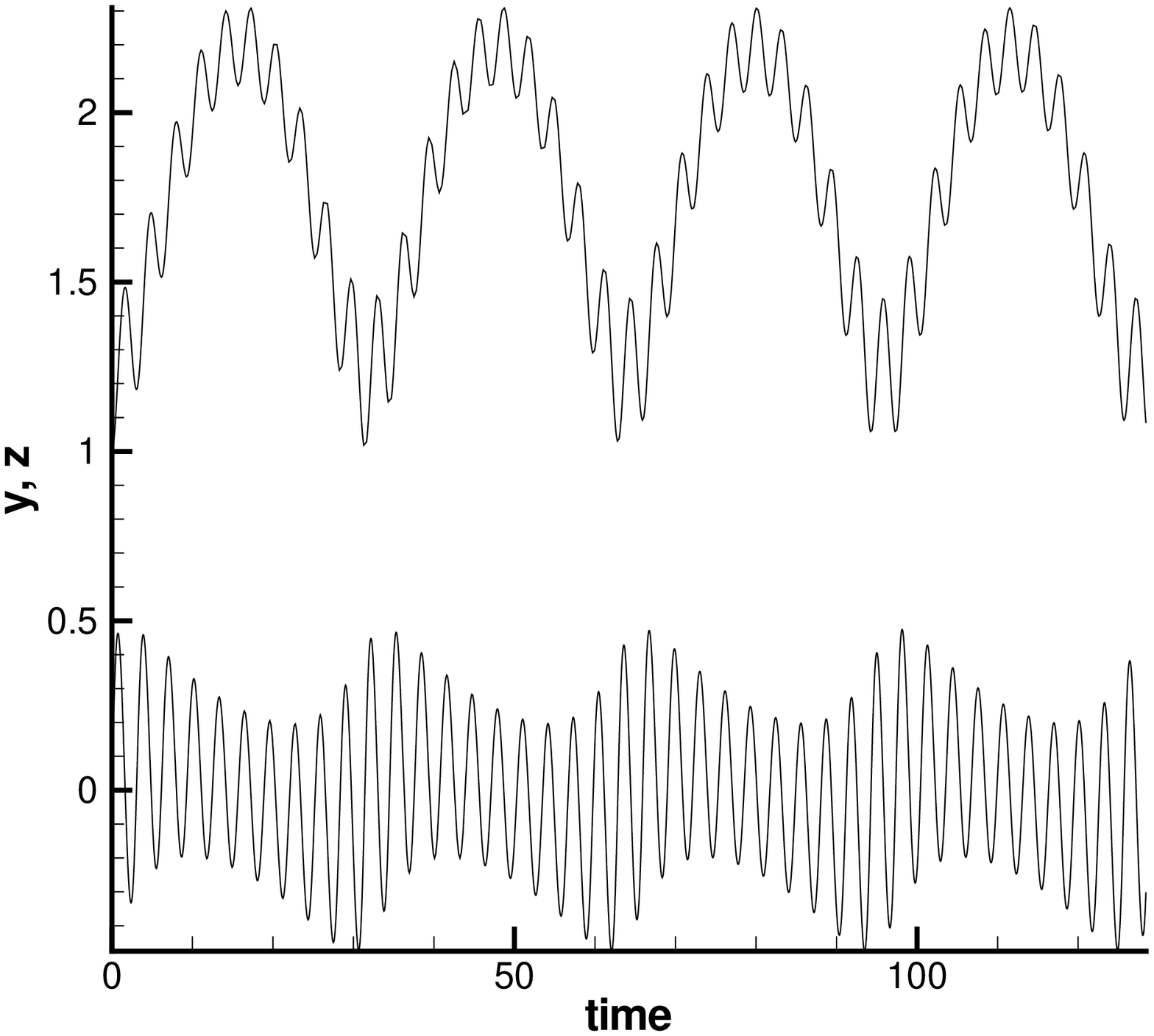}
\caption{Time dependence of non-dimensional radius $y$ (upper
curve), and non-dimensional velocity $z$ (lower curve), for
$D=2.11$.} \label{fig3}
\end{figure}


\begin{figure}
\includegraphics[angle=-90,width=0.45 \textwidth]{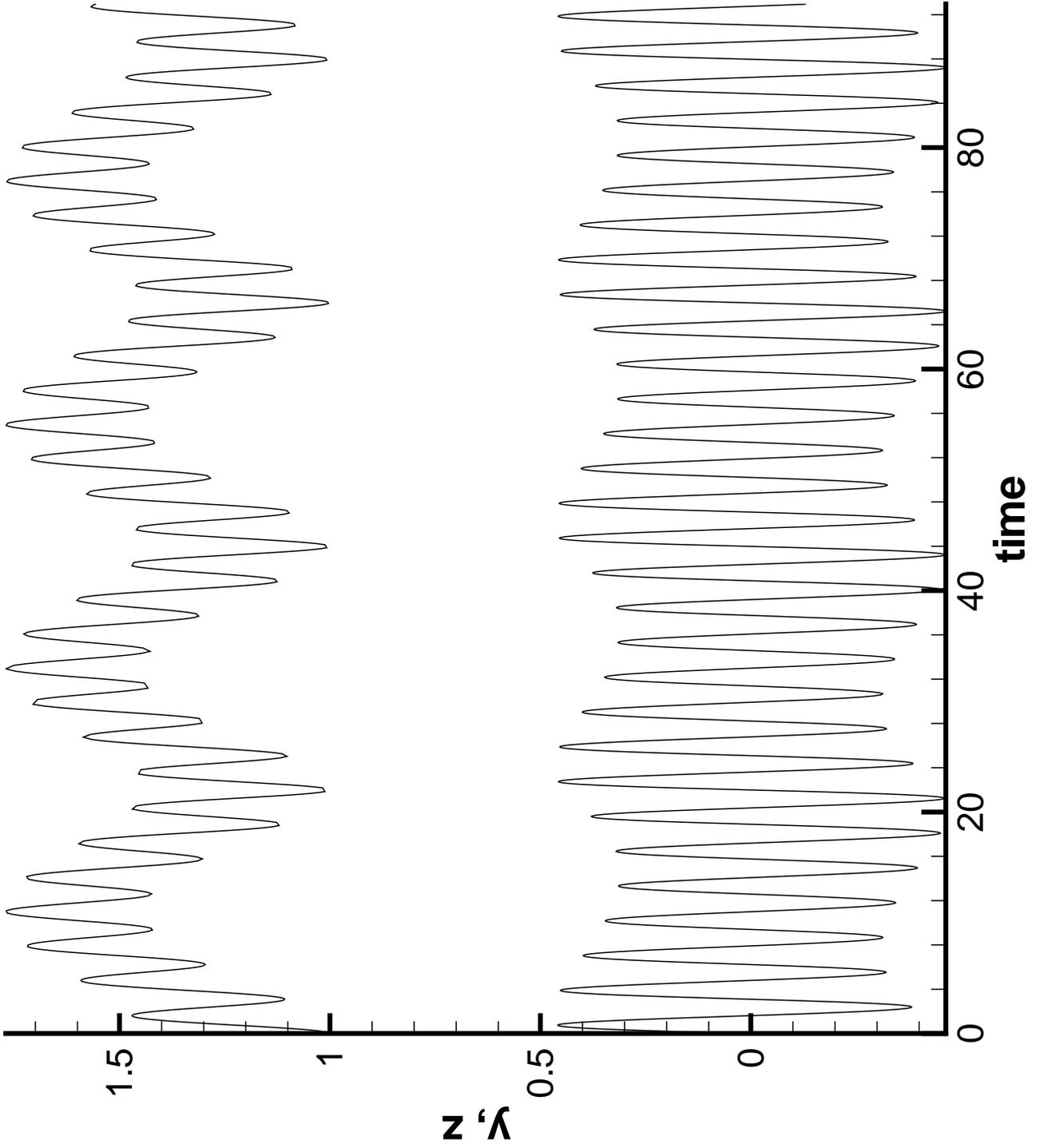}
\caption{Time dependence of non-dimensional radius $y$ (upper
curve), and non-dimensional velocity $z$ (lower curve), for $D=2.15$
during a long time period.} \label{fig3a}
\end{figure}

\begin{figure}
\includegraphics[angle=-90,width=0.45 \textwidth]{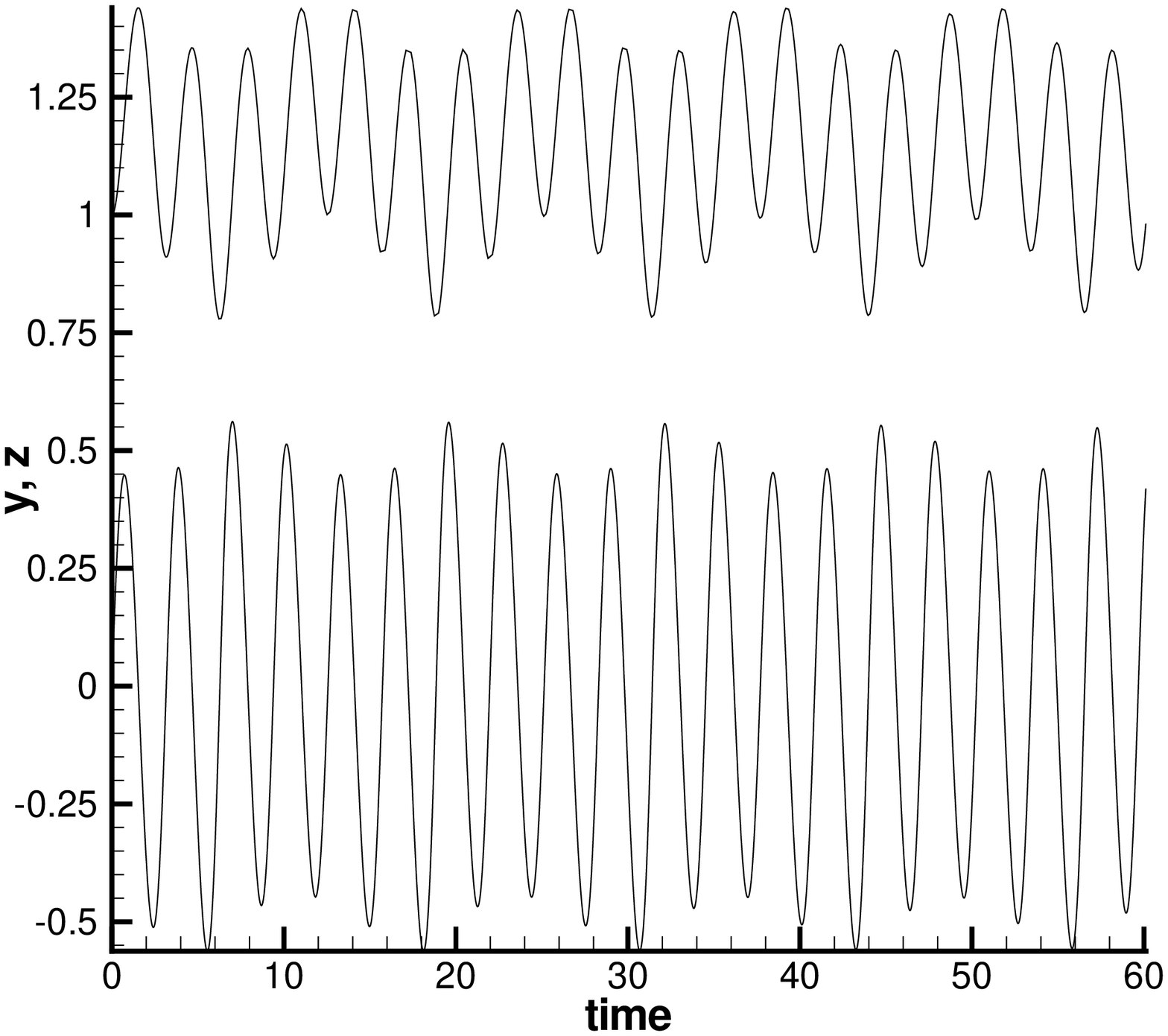}
\caption{Time dependence of non-dimensional radius $y$ (upper
curve), and non-dimensional velocity $z$ (lower curve), for
$D=2.25$.} \label{fig8}
\end{figure}

\begin{figure}
\includegraphics[angle=-90,width=0.45 \textwidth]{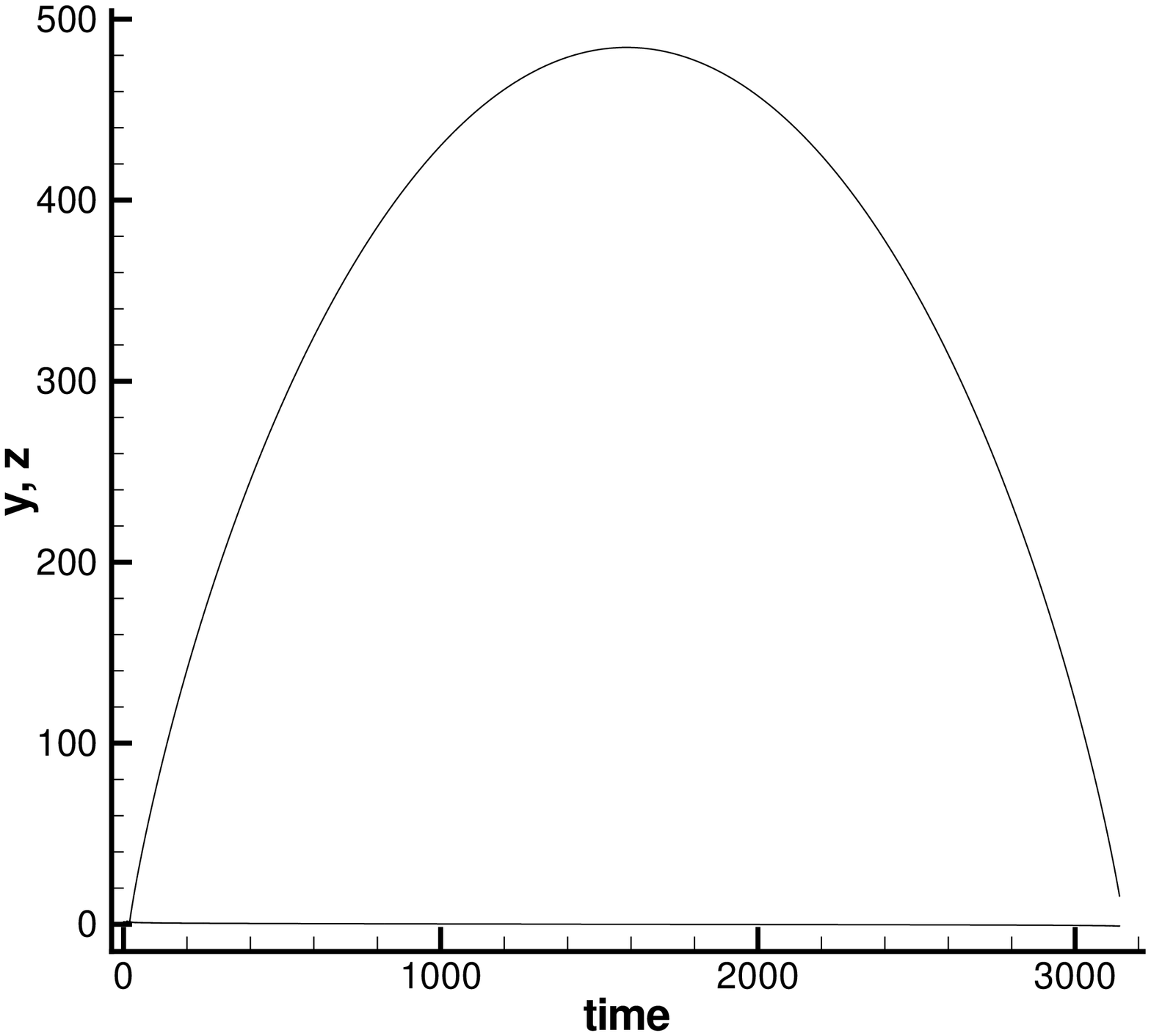}
\caption{Time dependence of non-dimensional radius $y$ (upper
curve), and non-dimensional velocity $z$ (lower curve), for
$D=2.3$.} \label{fig10}
\end{figure}

Introduce non-dimensional variables
$\tau=\omega t, \,\, y=\frac{\tilde R}{R_0},\,\,
z=\frac{a \tilde R}{a_0 R_0},\,\,
 a_0=\frac{K}{\omega R_0^2}=\omega,\,\,R_0=\frac{\sqrt K}{\omega},
$
in which differential equations have a form
\begin{equation}
\label{eq54} \frac{dy}{d\tau}=z,\, \frac{d
z}{d\tau}=\frac{1}{y}(1-D\sin^2 \tau),\,
y(0)=1,\,
z=0\,{\rm at}\, \tau=0.
\end{equation}
The problem is reduced to a system (\ref{eq54}) with
two non-dimensional parameters: $D=\frac{1}{2\pi K
C_m}\left(\frac{C_b\Omega_0}{z_0\omega}\right)^2$, and $y(0)$, and
the second one is taken equal to unity in farther consideration.
Solution of this nonlinear system
changes qualitatively with changing of the parameter $D$.

The solution of this system obtained numerically for different $D$ in the
interval between 1.5 and 3.2, may be divided into 3 groups.

1. At $D \le 2$ there is no confinement, radius grows to
infinity after several low-amplitude oscillations (Fig.2).

2. With growing of $D$ the amplitude of oscillations increase, and
at $D=2.1$ radius is not growing to infinity, but is oscillating
around some average value, forming rather complicated curves (Figs.
3-5).

3. At $D \ge 2.28$ the radius goes to zero. At $D=
2.28 -- 2.9$ the dependence of the radius $y$ with time may be very
complicated, consisting of low-amplitude and large-amplitude
oscillations, which finally lead to zero. The time at which radius
becomes zero may happen
at $\tau \le 100$, like at $D$=2.4, 2.6 (Bisnovatyi-Kogan, 2006), or goes trough
very large radius, and returned back to zero value at very large
time $\tau \sim 10^7$ at $D$=2.3, 2.55, 2.8 (Figs. 6-9). At $D\ge 3$
the radius goes to zero at
$\tau<2.5$ (Fig. 10), before the right side of the second equation
(\ref{eq54}) returned to the positive value.
\section{Discussion}
\label{sec:6}
\begin{figure}
\includegraphics[angle=-90,width=0.45 \textwidth]{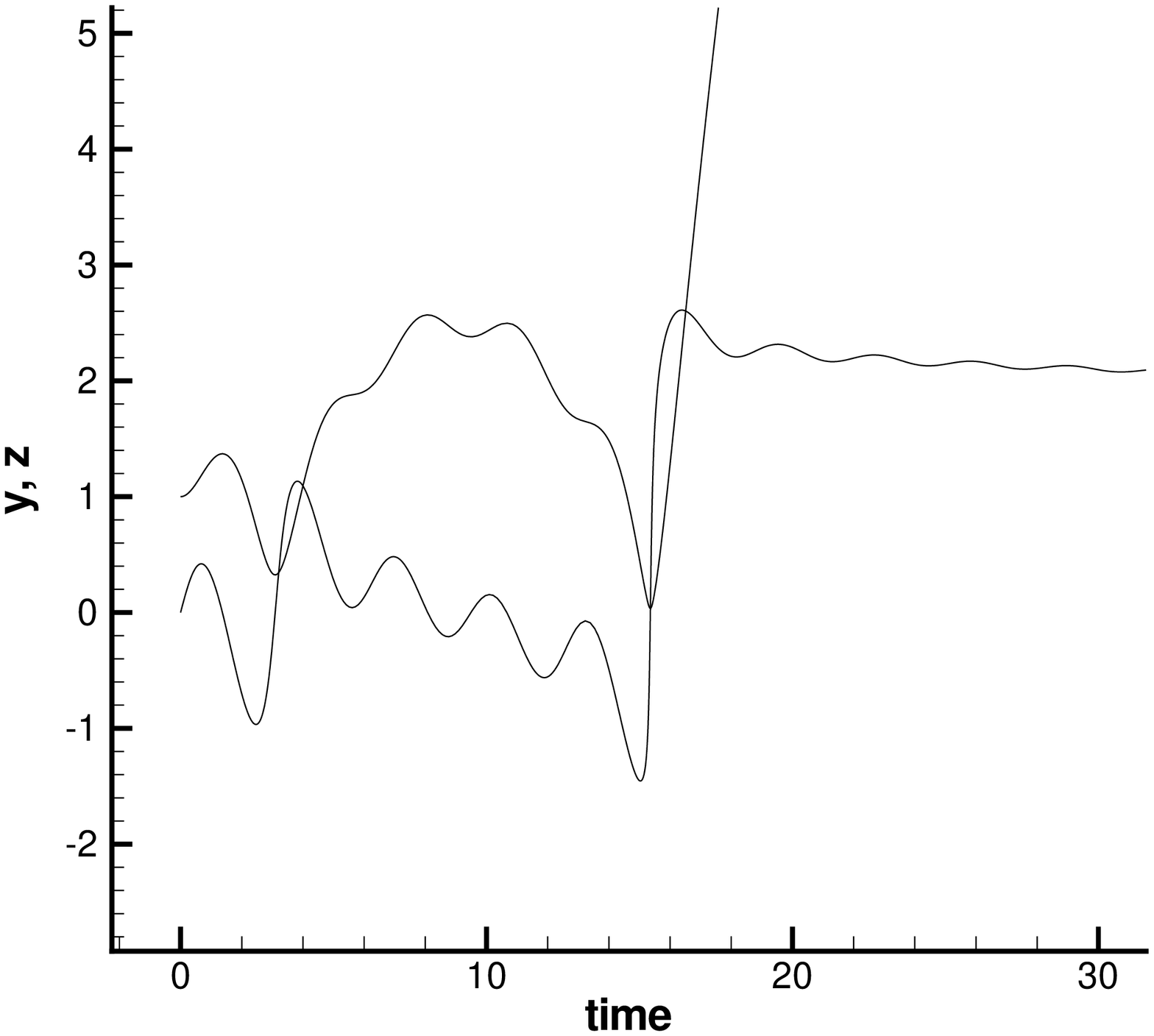}
\caption{Time dependence of non-dimensional radius $y$ (upper
curve), and non-dimensional velocity $z$ (lower curve), for
$D=2.55$.} \label{fig11}
\end{figure}

\begin{figure}
\includegraphics[angle=-90,width=0.45 \textwidth]{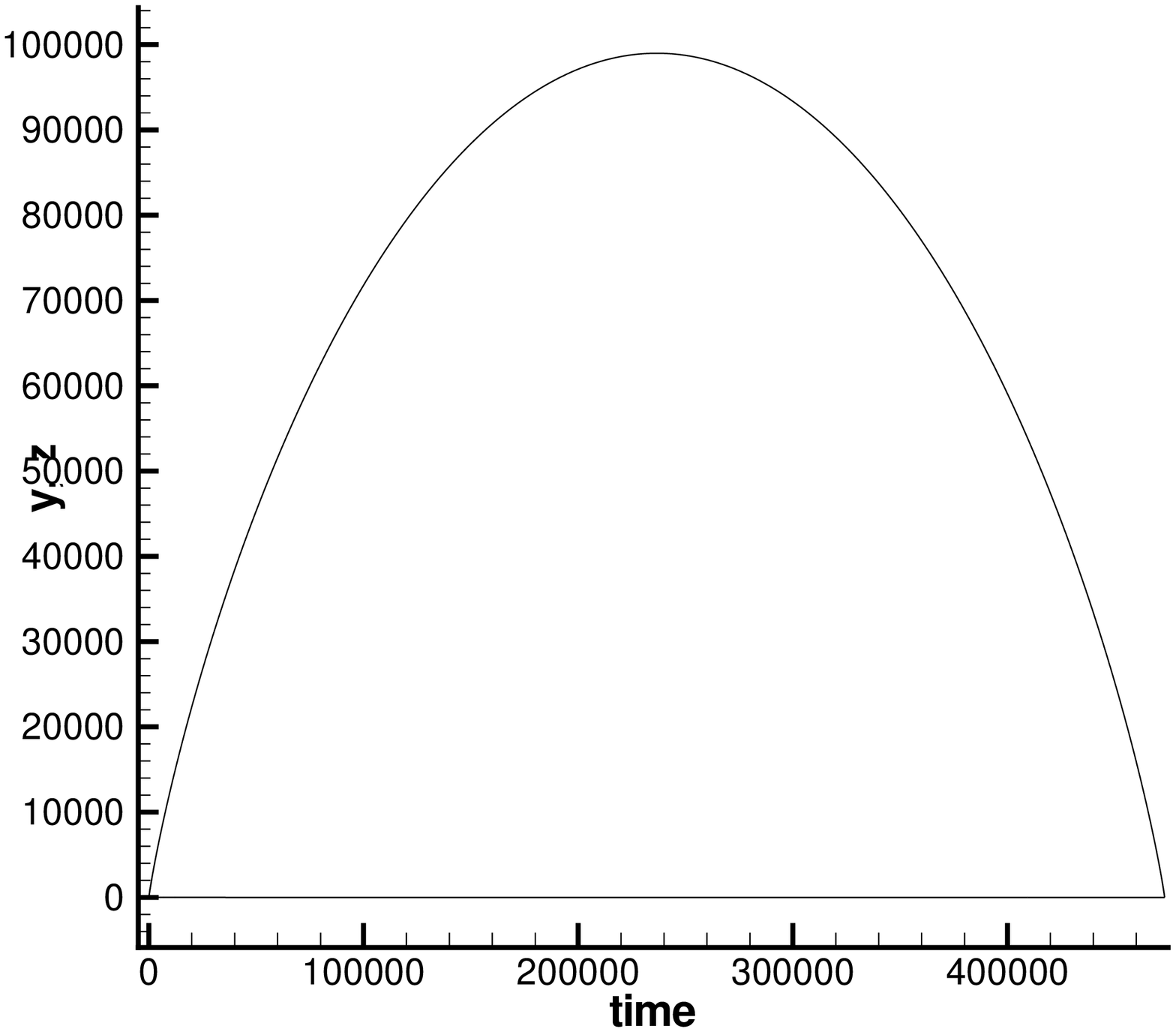}
\caption{Time dependence of non-dimensional radius $y$ (upper
curve), and non-dimensional velocity $z$ (lower curve), for
$D=2.55$, during a long time period.} \label{fig11a}
\end{figure}

\begin{figure}
\includegraphics[angle=-90,width=0.45 \textwidth]{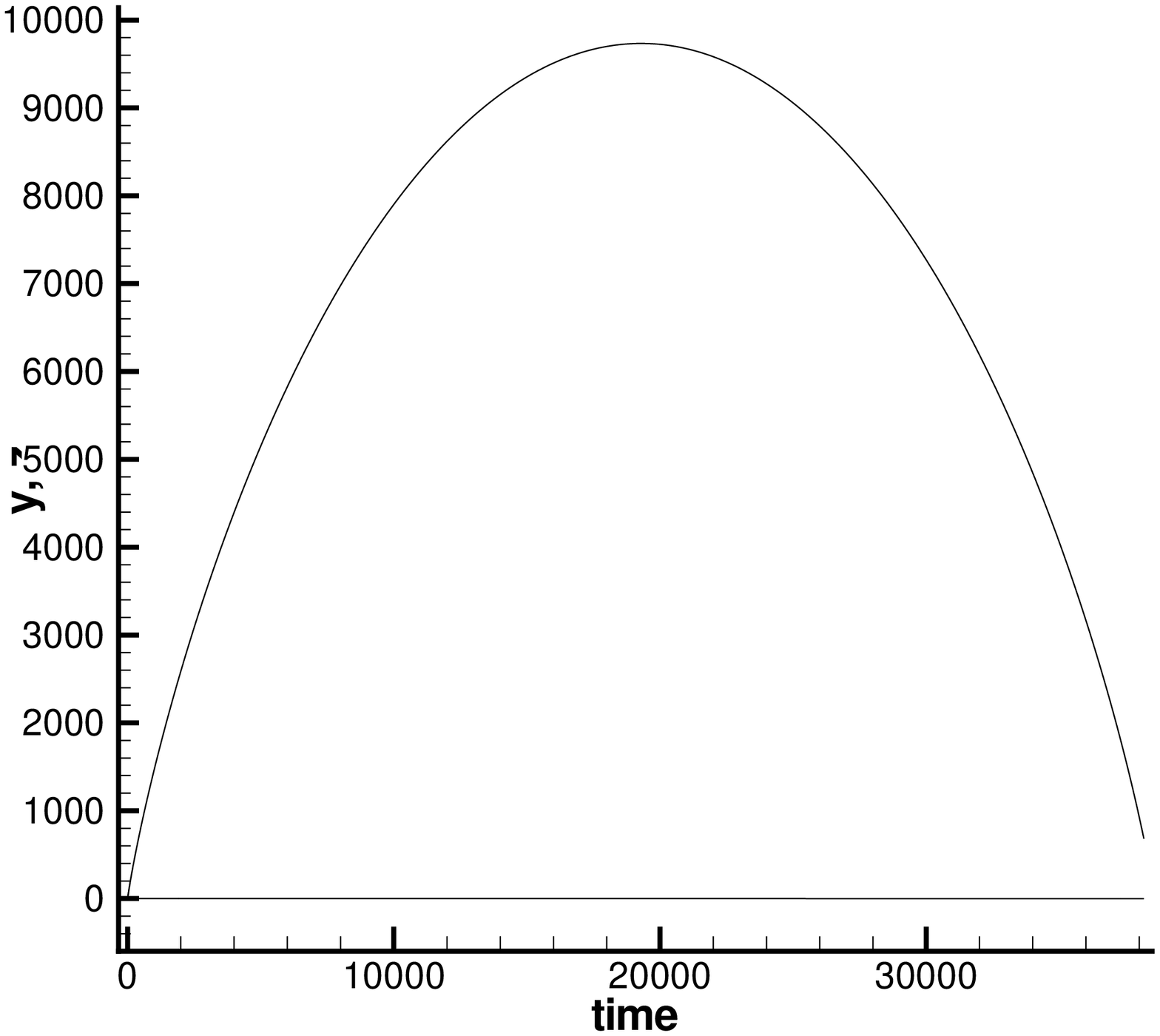}
\caption{Time dependence of non-dimensional radius $y$ (upper
curve), and non-dimensional velocity $z$ (lower curve), for
$D=2.8$.} \label{fig16}
\end{figure}

\begin{figure}
\includegraphics[angle=-90,width=0.45 \textwidth]{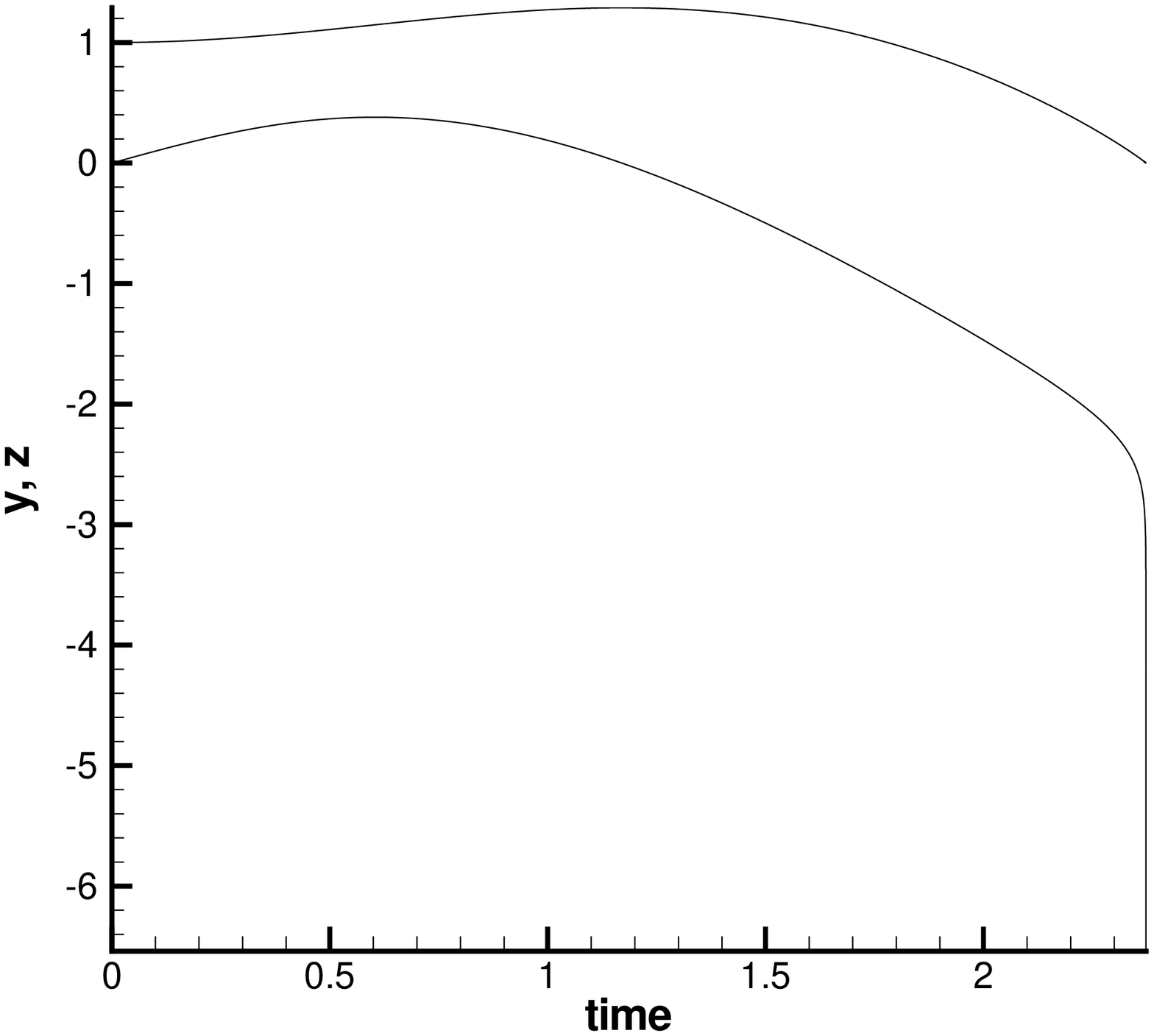}
\caption{Time dependence of non-dimensional radius $y$ (upper
curve), and non-dimensional velocity $z$ (lower curve), for $D=3.1$,
during a long time period.} \label{fig16a}
\end{figure}

Consider isothermal equation of state
$P=K\rho=v_s^2 \rho$, $v_s^2 \le c^2/3$, $v_s$ is the
sound speed. For ultrarelativistic pair-plasma
$K=c^2/3$. The parameter $D$, as a function of $R_0$, $z_0$,
initial $\rho_0$ and $B_{z0}$, $\Omega_0$
and $\omega$ is written as
\begin{equation}
\label{eq55} D=\frac{1}{2\pi\rho_0}\frac{B_{z0}^2 R_0^2
\Omega_0^2}{z_0^2 \omega^2 v_s^2}.
 \end{equation}
The amplitude of oscillations $\Omega$, and $\omega$ should be found
from the solution of the nonlinear system (\ref{eq19})-(\ref{eq26}),
together with the interval of values of $D$ at
which confinement happens. In the approximate system (\ref{eq54})
only $D$ characterizes different regimes, what for
given $R_0$, $z_0$, $\rho_0$, and $B_{z0}$ determines a
function $\Omega_0(D,\omega)$ for the collimated jet. To find
approximately a self-consistent model with $\Omega_0,\omega(D)$ we
use the the frequency linear oscillations
 $\omega=k V_A$. The
frequency of non-linear oscillations is smaller, so we may write
\begin{equation}
\label{eq56} \omega^2=\alpha_n^2\,k^2\,V_A^2\,=\,\alpha_n^2
\frac{\pi B_{z0}^2}{\rho_0 z_0^2},\,\, \alpha_n < 1, \,\,
k=\frac{2\pi}{z_0},
 \end{equation}
therefore
$
\Omega^2 R_0^2=2\pi^2 D \alpha_n^2 v_s^2 <c^2,\,\,
R_0^2=\frac{K}{\omega^2}=z_0^2\frac{\rho_0 v_s^2}{\alpha_n^2 \pi
B_{z0}^2}.
$
On the edge of the cylinder the rotational velocity cannot exceed
the light velocity, so the solution has a physical sense only
at $v_s^2 < \frac{c^2}{2\pi^2 D \alpha_n^2} \approx \frac{c^2}{40
\alpha_n^2}$. Taking $\alpha_n^2=0.1$ for non-linear
oscillations we obtain a restriction $v_{s0}^2 <
\frac{c^2}{4}$.
To have the sound velocity not exceeding $c/2$, the jet should
contain baryons, which density $\rho_0$ exceeding about 30\% of
 the total density of the jet.
In a dense quasispherical stellar cluster around a supermassive
black hole, the accretion disk is changing its direction of
rotation, due to different sign of the angular momentum of the
falling stars. In this sutuation the  magneto-torsional oscillations
should be inevitably generated in the outflowing jets. The knots
 in jets are observed in different objects, including a famous M87 jet.
 In the last one the knots are visible in radio,
optics (Hiltner, 1959), and X-rays (Wilson and Yang, 2002). It is
possible that they are connected with magneto-torsional
oscillations.

\medskip

{\bf Acknowledgement} This work was partially supported by RFBR
grants 05-02-17697, 06-02-91157, 06-02-90864, and President grant
for a support of leading scientific schools 10181.2006.2.

\end{document}